\documentclass[aip, apl, reprint]{revtex4-1}
\usepackage{graphics}
\usepackage{graphicx}
\usepackage{bm}
\usepackage[normalem]{ulem}

\begin{document}

\title{Spectroscopic measurements of a CO$_2$ absorption line in an open vertical path using an airborne lidar}

\author{Anand Ramanathan}
\email{anand.ramanathan@nasa.gov}
\author{Jianping Mao}
\affiliation{Earth System Science Interdisciplinary Center, University of Maryland College Park MD-20740 USA}

\author{Graham R. Allan}
\affiliation{Sigma Space Corporation, Lanham, MD-20706, USA}

\author{Haris Riris}
\affiliation{NASA Goddard Space Flight Center, Greenbelt, MD-20771, USA}

\author{Clark J. Weaver}
\affiliation{Earth System Science Interdisciplinary Center, University of Maryland College Park MD-20740 USA}

\author{William E. Hasselbrack}
\affiliation{Sigma Space Corporation, Lanham, MD-20706, USA}

\author{Edward V. Browell}
\affiliation{STARSS-II Affiliate, NASA Langley Research Center, Hampton, Virginia 23681 USA}

\author{James B. Abshire}
\affiliation{NASA Goddard Space Flight Center, Greenbelt, MD-20771, USA}

\date{\today}

\begin{abstract}

We used an airborne pulsed integrated path differential absorption (IPDA) lidar to make spectroscopic measurements of the pressure-induced line broadening and line center shift of atmospheric carbon dioxide (CO$_2$) at the 1572.335 nm absorption line. We scanned the lidar wavelength over 13 GHz (110 pm) and measured the absorption lineshape at 30 discrete wavelengths in the vertical column between the aircraft and ground. A comparison of our measured absorption lineshape to calculations based on HITRAN shows excellent agreement with the peak optical depth accurate to within 0.3\%. Additionally, we measure changes in the line center position to within 5.2 MHz of calculations, and the absorption linewidth to within 0.6\% of calculations. These measurements highlight the high precision of our technique, which can be applied to suitable absorption lines of any atmospheric gas.  
\end{abstract}

\maketitle

Optical remote sensing of trace gases in the atmosphere is widely used to understand terrestrial processes\cite{MOPPITT2003CO, SatelliteMethane2005, SatelliteNOx2006, SatelliteReview2013}. However, the accuracy of such remote sensing measurements can be limited by the accuracy of the spectroscopy of the absorption lines\cite{InfluenceLM2009CO2}. Lab-based precision laser spectroscopy measurements are difficult to validate in the open atmosphere with passive remote sensing instruments\cite{SatelliteReview2013} owing to complications arising from clouds and aeorosols\cite{SCIAMACHY2005Errors, GOSAT2012Aerosols} and uncertainties in the air mass factor\cite{OMI2011AirMassFactor}. In this letter, we report laser absorption spectroscopy measurements of carbon dioxide (CO$_2$) made in an open vertical atmospheric path. Using an aircraft-mounted pulsed lidar, we study the effect of pressure on the 1572.335 nm (6359.967 cm$^{-1}$) line of the CO$_2$ vibrational-rotational band. 

The absorption spectra of atmospheric CO$_2$ has been studied by satellite\cite{SCIAMACHY2011CO2, GOSAT2011} and ground-based passive spectrometers\cite{TCCON}, as a means to obtain the CO$_2$ distribution. IPDA measurements from aircraft have also been demonstrated\cite{Spiers2011, Sakaizawa2013, Dobler2013, abshire2013} as part of efforts to develop space-borne lidars. Converting absorption measurements to CO$_2$ concentrations requires precise spectroscopic knowledge, which is typically derived from controlled laboratory experiments\cite{malathy2007line, sakaizawa2008PressureShift, long2011air, christensen2012tunable}. Our present work extends spectroscopic investigations to conditions closer to those of atmospheric remote sensing.

\begin{figure}[!ht]
\includegraphics[width=\columnwidth , clip=true, trim = 0mm 0mm 0mm 0mm]{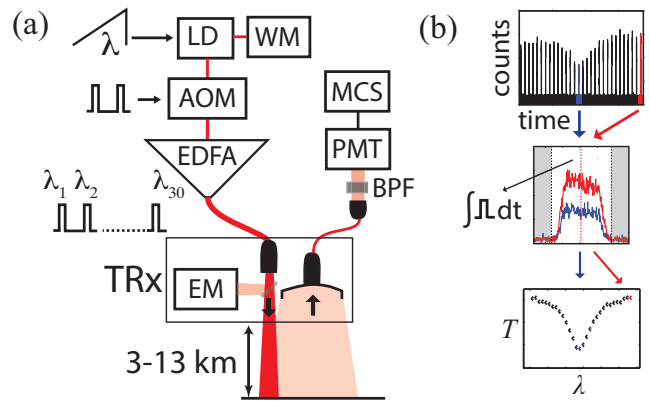}
  \caption{(Color online) Airborne CO$_2$ sounder lidar instrument schematic- (a) A repeating 30-wavelength lidar pulse train is generated using a gated diode laser (LD), which are then amplified using an Erbium-doped fiber amplifier (EDFA). The ground returns are detected using a photo-multiplier tube (PMT) and accumulated on a multi-channel scaler (MCS). Note: AOM=acousto-optic modulator, WM=wavelength monitor, EM=energy monitor, BPF=bandpass filter to block the solar background, TRx-Transceiver. (b) The top diagram shows the return pulse train with highlighted on-line and off-line pulses (middle). We integrate across individual pulses to obtain the transmittance vs wavelength (bottom) for the absorption line.}
  \label{fig:Plane}
\end{figure}

The lidar instrument, primarily designed for measuring the tropospheric CO$_2$ concentration in the column beneath the aircraft\cite{abshire2010, abshire2013}, was flown aboard the NASA DC-8 during the ASCENDS\cite{ASCENDSWG} (Active Sensing of CO$_2$ Emissions over Nights, Days and Seasons) campaign in 2011, at altitudes ranging from 2 to 13 km above sea level. The 1572.335 nm absorption line was chosen based on its optimal line strength and temperature sensitivity\cite{Mao2004}.  A schematic of our lidar instrument is shown in FIG.~\ref{fig:Plane}(a) and the relevant system parameters tabulated (table~\ref{tab:parameters}). Our lidar source was a distributed feedback diode laser amplified by an erbium-doped fiber amplifier (EDFA) operating in a master oscillator power amplifier configuration. The laser was gated with an acousto-optic modulator (AOM) to transmit a train of discrete pulses. This pulse scheme allows for isolating cloud, aerosol and ground returns, enables precision ranging to the reflecting surface and ensures only one lidar pulse in the atmosphere. The outgoing energy of individual pulses (``EM" in FIG.~\ref{fig:Plane}a) was monitored using a weak beamsplitter, integrating sphere and detector. The outgoing lidar beam was pointed nadir from the aircraft.

In addition to gating, the source diode laser wavelength was continuously scanned\cite{abshire2010} (by changing the laser driver current) at $\approx$300 Hz so as to obtain 30 equally-spaced discrete wavelengths that were repeatedly transmitted. During operation, the wavelength scan shape stays constant, except for a wavelength offset that may drift over the course of hours. Our technique does not require maintaining absolute wavelength stability but rather stable separation intervals between wavelengths. We pre-calibrated the wavelength scan (prior to the AOM gating) using optical heterodyne detection\cite{OpticalHeterodyne1992}. During flight, we monitored the laser wavelength using a CO$_2$ absorption cell (``WM" in FIG.~\ref{fig:Plane}a) and a wavemeter.  

\begin{table}
\begin{tabular}{ l r}
\hline
Parameter & Value \\
\hline
lidar center wavelength & 1572.335 nm \\
no. of wavelengths sampled & 30 \\
wavelength spacing & $\approx$4 pm (500 MHz)\\
wavelength span & $\approx$110 pm (13 GHz) \\
wavelength scan repetition rate & $\approx$300 Hz\\
pulse duration & 1 $\mu$s \\
pulse energy & 20 $\mu$J \\
pulse repetition rate & 10 kHz (100 $\mu$s sep.) \\
lidar beam divergence & 100 $\mu$rad \\
lidar spot size on ground & $\sim$1 m \\
aircraft speed & $\approx$250 m/s \\
receiver telescope diameter & 20 cm \\
receiver field-of-view & 200 $\mu$rad \\
receiver filter bandpass & $\sim$0.8 nm (100 GHz)\\ 
receiver time resolution & 8 ns \\
data averaging time & 50 s\\
\hline
\end{tabular}
\caption{Airborne lidar instrument parameters}
\label{tab:parameters}
\end{table}

Our receiver consisted of a 20 cm diameter telescope that captured the return light, which was then fiber coupled and sent to a near IR photo-multiplier tube (PMT). For sensitive detection, the PMT was used in a high-gain photon counting mode. A discriminator converted the PMT output to discrete pulses that indicate photon counts. These pulses were then sent to a multi-channel scaler (MCS), which counted the pulses and binned them by arrival time. The MCS was synchronized to the wavelength scan. The MCS accumulated photon counts over $\approx$300 wavelength scans ($\approx$900 ms) before the data was retrieved by the control computer. The instrument recorded data in 1 s intervals (FIG.~\ref{fig:Plane}b top). 

\begin{figure}
\includegraphics[width=\columnwidth, clip=true, trim = 0mm 3mm 0mm 21mm]{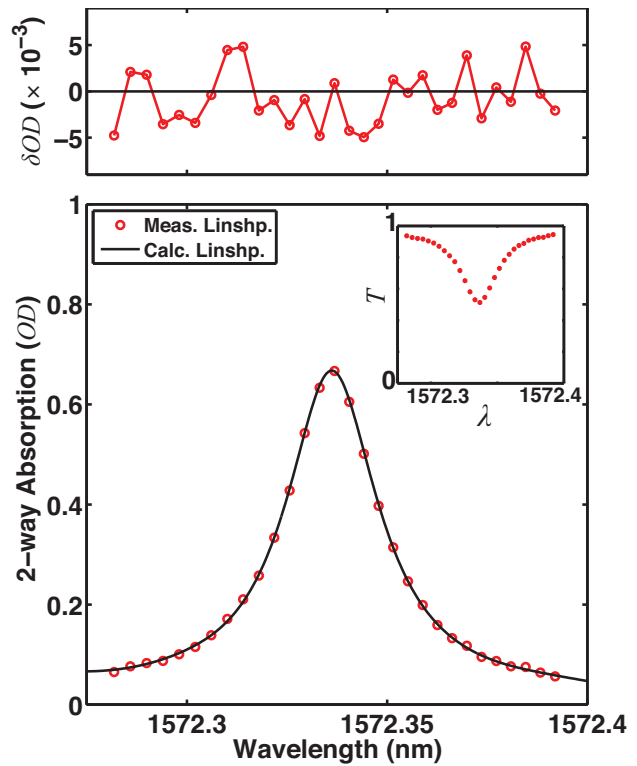}
  \caption{(Color online) Lidar lineshape measurement - (bottom) We express the atmospheric column's transmittance measurement (inset) as an absorption lineshape (circles) and compare it to LUT calculations (line). The residual (top) is small with $\delta OD_\textrm{rms}$ = 2$\times$10$^{-3}$ ($\approx$0.3\% at peak). Note: aircraft altitude=4.5 km, averaging time=50 s.}
  \label{fig:SampleLineshape}
\end{figure} 

The measured transmission lineshape as well as the optical range are obtained from the MCS data. We integrate over the individual pulses (illustrated in FIG.~\ref{fig:Plane}(b)) after subtracting the background (solar background and PMT dark current) to obtain the respective individual return pulse energies. The pulse time-of-flight gives the optical range\cite{Amediek2013}, the one-way distance traveled by the lidar beam (accurate to 3 m). For the results shown in this letter, we additionally average the 1 s pulse energies over 50 s intervals to reduce photon shot noise. The time-averaged pulse energies are first normalized by the energy monitor measurements and then by the mean energy of the off-peak pulses (we choose pulses 2-5 and 26-30) to remove any dependence on varying outgoing laser energy and surface reflectivity respectively. We also allow for a small, linear baseline system wavelength response and correct for it\cite{abshire2013}. Combining the pulse energies with the onboard wavelength calibration, we obtain the atmospheric transmission lineshape (FIG.~\ref{fig:SampleLineshape} inset). We then compute the optical depth (FIG.~\ref{fig:SampleLineshape} circles), $OD (\lambda)=-\ln T(\lambda)$, where $T$ is the transmission. We use this absorption lineshape for further analysis.

\begin{figure}
\includegraphics[width=\columnwidth]{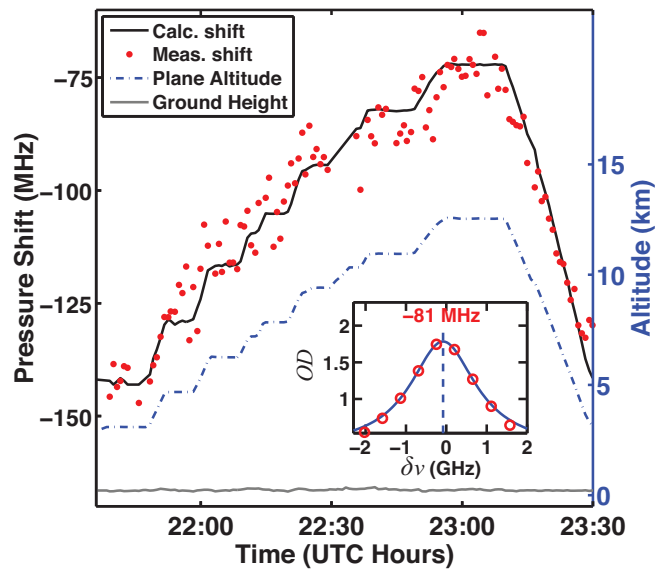}
  \caption{(Color online) Altitude-dependent pressure shift: The pressure shift (red dots) is obtained by fitting a Lorentzian to determine the center of the absorption lineshape (inset - $\delta\nu$= frequency relative to natural line center\cite{HITRAN2008}). These measurements agree to within 5.2 MHz (rms difference) of calculations (black solid line) computed from the LUT factoring the flight altitude (blue dash-dot, right axis) and ground elevation (gray solid).}
  \label{fig:PressureShift}
\end{figure}

The measured absorption lineshape depends on several factors such as the atmosphere  pressure, temperature, water vapor and CO$_2$ concentration profile. We get this information by flying the aircraft (and instrument) in a particular pattern and making some simplifying assumptions regarding the atmosphere. The data shown in this letter was taken from a flight segment over Iowa on August 10$^{\textrm{th}}$ 2011, where we flew the aircraft back and forth over the same ground track at altitudes from 3 to 12 km (see FIG.~\ref{fig:PressureShift} right axis), followed by a descent spiral from 12 km to the ground over the center of the track.

During the descent spiral, \textit{in situ} instruments collected vertically resolved atmospheric pressure, temperature and water vapor information (DC-8 aircraft instrumentation) as well as the CO$_2$ concentration profile (on board AVOCET\cite{AVOCET}). Post flight, these data were used to create a look-up-table (LUT) of monochromatic optical depths (at wavelengths 1572.0-1572.5 nm) for 50 m layers from the surface to the highest flight altitudes (12 km). In creating the LUT, we used the Line-By-Line Radiative Transfer Model\cite{Clough1992, Clough2005} (LBLRTM) and HITRAN 2008\cite{HITRAN2008} to calculate CO$_2$ and H$_2$O absorptions. CO$_2$ line-mixing\cite{Lamouroux2010} was included in the calculations. The LUT is used in combination with the GPS position, aircraft attitude and lidar range to calculate the model lineshape (black line in FIG.~\ref{fig:SampleLineshape}) for comparison with the lidar data.  

Our data agrees well with the calculations as seen in FIG.~\ref{fig:SampleLineshape}. We quantify this comparison by subtracting the calculated lineshape from the measured lineshape and plotting the residuals (shown in FIG.~\ref{fig:SampleLineshape} top). The rms error in OD, 2$\times$10$^{-3}$ is small compared the range of measured OD values ($\approx$0.6) across the lineshape. To assess the measurement precision of our technique, we extract two lineshape properties, linewidth and line center and compare them to our atmospheric model calculations based on the LUT. Both lineshape characteristics depend primarily on the atmospheric pressure profile and to a lesser extent, on the temperature profile.

The atmospheric pressure decreases with altitude, from about 1010 mbar at the ground level to about 190 mbar at the highest altitude flown. Higher pressures causes more broadening of the absorption line at low altitudes. In addition, higher pressures cause a small redshift in the absorption line center, the pressure shift\cite{PressureShift}. The measured absorption linewidth and line center position depends on the combined effect of the atmospheric pressure profile from the ground to the plane's altitude. The CO$_2$ 1572 nm half-width at half-maximum (HWHM) linewidth of the total column absorption ranges from $\approx$2 GHz (17 pm) at low altitudes to $\approx$1 GHz (8 pm) at the highest altitudes flown. Similarly, the total column pressure shift\cite{sakaizawa2008PressureShift} of the line center ranges from $\approx$-150 MHz (1.3 pm) to $\approx$-70 MHz (0.6 pm). 

Since the peak absorption may lie between adjacent lidar wavelengths, we need to interpolate between wavelengths to determine the precise location of the line center. However, given the effects of line-mixing, thermal doppler broadening and the pressure profile of the atmosphere, there is no simple function to express the absorption lineshape. An effective technique to determine the line center is to fit a Lorentzian function to the center 5-10 wavelengths of the absorption lineshape (FIG.~\ref{fig:PressureShift} inset) to obtain the peak absorption and line center. Using a Lorentzian function (as opposed to our model calculations) to extract the line center ensures that our measurement is independent of the LUT. 

Comparing the measured pressure shift with calculations (FIG.~\ref{fig:PressureShift}) showed that they agree within $\approx$15 MHz (0.12 pm) across the entire range of flight altitudes. While not critical, in factoring Doppler shifts\cite{Doppler} that arise from the relative velocity between the aircraft (lidar instrument) and air mass, the accuracy of our measurement improved by 20\%. The rms difference between the measured pressure shift and calculations is 5.2 MHz (0.043 pm), which is significantly smaller than the 500 MHz (4 pm) wavelength spacing and 13 GHz (110 pm) span. The measurement accuracy with the present lidar instrument is limited by variations in surface reflectivity.

\begin{figure}
\includegraphics[width=\columnwidth]{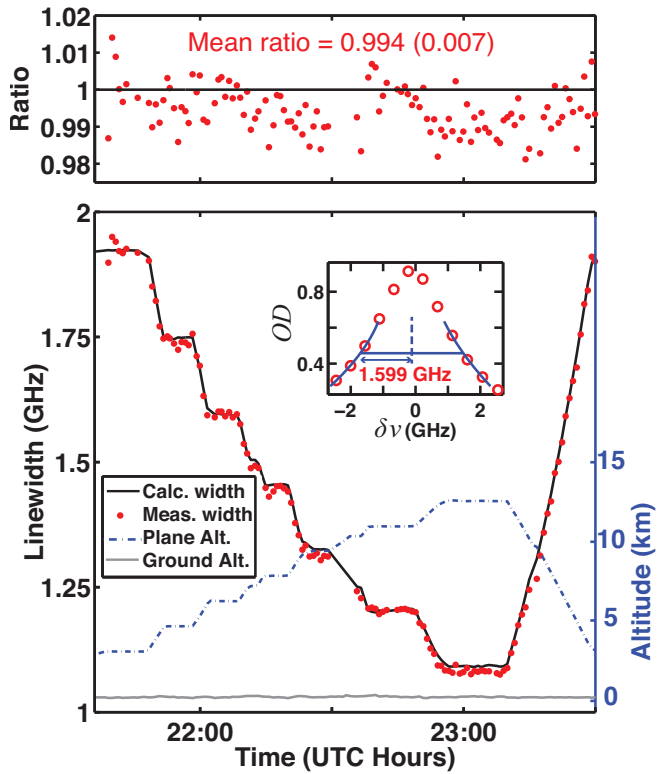}
  \caption{(Color online) Altitude dependent pressure broadening: The half-maximum linewidth (bottom panel, red dots) is interpolated from the absorption lineshape (inset). These measurements are in excellent agreement with the linewidth computed from the LUT (black solid line) factoring the flight altitude (blue dash-dot, right axis) and ground elevation (gray solid). The mean ratio of the measured linewidth to calculations (top panel) is within 0.6\% of unity (standard deviation = 0.7\%).}
  \label{fig:PressureBroadening}
\end{figure}

The linewidth (HWHM) is determined from the location of the two half-maximum positions. Again, since these precise points may not coincide with a wavelength that we sample, we quadratically interpolate the points from the data (FIG.~\ref{fig:PressureBroadening} inset). The measured linewidth (shown in FIG.~\ref{fig:PressureBroadening}) is within 0.6\% of calculations (from the LUT), limited primarily by our knowledge of the vertical distribution of CO$_2$. This agreement highlights the high precision with which our technique can make open path spectroscopic measurements.

Although the present measurements of the line center and linewidth may not provide new knowledge about the physics of lineshapes, they constitute validation of the LBLRTM\cite{Clough2005} and HITRAN\cite{HITRAN2008} in nature across a wide range of temperatures and pressures. This comparison is possible thanks to the stability of the instrument, the robust retrieval approach and the successful modeling of the atmosphere in terms of the LUT.

In summary, we have made spectroscopic measurements of a CO$_2$ line in the open atmosphere along nadir paths of 3-12 km. These measurements were made from an airborne platform to rough and varied terrain in the presence of winds, clouds and aerosols. Our present lidar resolves the line center to within 5.2 MHz and the linewidth to within 0.6\% ($\approx$10 MHz). As a result, we observe small effects from phenomena such as Doppler shifts due to winds and aircraft velocity. Tuning the wavelength sampling to accurately probe specific CO$_2$ spectroscopic features may allow for remote measurements of wind, pressure, temperature\cite{LAS2006Temp} or other atmospheric constituent gas concentration\cite{LAS2013GasConc} using lineshape information. 

Our technique, combined with the appropriate lasers and optics, can be used to provide open path spectroscopic measurements for satellite-based remote sensing\cite{SatelliteReview2013} of CO$_2$, methane, oxides of nitrogen and ozone. Further, by coordinating airborne lidar measurement with a satellite overpass, one can cross-check satellite measurements and help understand any deviations from calculations. Thus, our technique broadens the scope of open path lidar measurements and can provide an important bridge that connects laser spectroscopy to satellite remote sensing of trace gases.

\acknowledgments     
 
This work was funded by the NASA ESTO IIP-10 program and the NASA ASCENDS definition program. AR would like to acknowledge extensive support from the NASA Postdoctoral Program. We also thank the AVOCET team of NASA LaRC for providing the \textit{in situ} CO$_2$ concentration measurements and the NASA DAOF DC-8 team for help conducting the flight campaign.


\begin{thebibliography}{33}
\bibitem [{Dop()}]{Doppler}%
  \BibitemOpen
  \href@noop {} {}\bibinfo {note} {Since the lidar is pointed towards nadir,
  noticeable doppler shifts occur only when the aircraft has a significant
  pitch or roll angle such as during climbs, descents or turns.}\BibitemShut
  {Stop}%
\end{thebibliography}
\end{document}